\documentclass[pdflatex,sn-basic]{sn-jnl}

\usepackage{graphicx}%
\usepackage{multirow}%
\usepackage{amsmath,amssymb,amsfonts}%
\usepackage{amsthm}%
\usepackage{mathrsfs}%
\usepackage[title]{appendix}%
\usepackage{xcolor}%
\usepackage{textcomp}%
\usepackage{manyfoot}%
\usepackage{booktabs}%
\usepackage{algorithm}%
\usepackage{algorithmicx}%
\usepackage{algpseudocode}%
\usepackage{listings}%

\theoremstyle{plain}
\newtheorem{theorem}{Theorem}
\theoremstyle{definition}
\newtheorem{proposition}{Proposition}
\newtheorem{lemma}{Lemma}

\usepackage{dsfont}
\usepackage{enumitem}
\usepackage{cleveref}
\usepackage{subcaption}
\usepackage{xcolor}

\counterwithin{figure}{section}
\numberwithin{equation}{section}
\crefname{equation}{Eqn.}{Eqns.}
\crefname{figure}{Fig.}{Fig.}

\begin{document}

\title[Article Title]{Transformations to simplify phylogenetic networks}


\author[1,2]{\fnm{Johanna} \sur{Heiss}}

\author[1]{\fnm{Daniel H.} \sur{Huson}}

\author[2]{\fnm{Mike} \sur{Steel}}\email{mike.steel@canterbury.ac.nz}

\affil[1]{\orgdiv{Institute for Bioinformatics and Medical Informatics}, \orgname{University of Tübingen}, \city{Tübingen}, \country{Germany}}

\affil[2]{\orgdiv{Biomathematics Research Centre}, \orgname{University of Canterbury}, \city{Christchurch}, \country{New Zealand}}

\abstract{The evolutionary relationships between species are typically represented in the biological literature by rooted phylogenetic trees. However, a tree fails to capture ancestral reticulate processes, such as the formation of hybrid species or lateral gene transfer events between lineages, and so the history of life is more accurately described by a rooted phylogenetic network. Nevertheless, phylogenetic networks may be complex and difficult to interpret, so biologists sometimes prefer a tree that summarises the central tree-like trend of evolution. In this paper, we formally investigate methods for transforming an arbitrary phylogenetic  network into a tree (on the same set of leaves) and ask which ones (if any) satisfy a simple consistency condition. This consistency condition states that if we add additional species into a phylogenetic network (without otherwise changing this original network) then transforming this enlarged network into a rooted phylogenetic tree induces the same tree on the original set of species as transforming the original network. We show that the LSA (lowest stable ancestor) tree method satisfies this consistency property, whereas several other commonly used methods (and a new one we introduce) do not. We also briefly consider transformations that convert arbitrary phylogenetic networks to another simpler class, namely normal networks.}

\keywords{phylogenetic networks, trees, transformations, lowest stable ancestor}

\maketitle

\newpage

\section{Introduction}\label{sec1}

The traditional representation of evolutionary history is based on rooted phylogenetic trees. Such trees provide a simple illustration of speciation events and ancestry. The leaves correspond to known species, and the root represents the most recent common ancestor of this set of species. However, a tree fails to capture ancestral reticulate processes, such as hybridisation events or lateral gene transfer. Thus, the evolutionary history is more accurately described by rooted phylogenetic networks~\citep{Huson2010, Steel2016}. Nevertheless, it is often helpful to extract the overall tree-like pattern that is present in a complex and highly reticulated network, sometimes referred to as  the `central tree-like trend' in the evolution of the taxa~\citep{Puigbo2013, Wolf2002}. Such a tree is less complete than a network, but it is  more easily interpreted and visualised~\citep{DeSalle2020}.

In this paper, we investigate ways to transform arbitrary rooted phylogenetic networks into rooted phylogenetic trees. There are many ways to do this, and we take an axiomatic approach, listing three desirable properties for such a transformation. We show that several current methods fail to satisfy all three properties; however, one transformation (the LSA tree construction) satisfies all three. Our approach is similar in spirit to an analogous axiomatic treatment of consensus methods (which transform an arbitrary set of trees into a single tree) in~\citet{Bryant2017}. We also briefly consider transformations that convert phylogenetic networks to `normal' networks (a class of networks that allows a limited degree of reticulation). We begin by defining some concepts that  will play a central role in our axiomatic approach.

\subsection{Preliminaries}

In this paper, we only consider rooted phylogenetic networks and trees on any leaf set $X$ of taxa. Such trees and networks are defined by a connected acyclic graph containing a set of vertices $V$ and a set of arcs $A$. We refer to arcs as edges, which are directed. The vertex $\rho$ is determined as the {\em root vertex} with an out-degree of at least $2$, where all edges are directed away from $\rho$. The set of labelled leaves is defined by $X \subseteq V$, which are vertices of an in-degree $1$ and an out-degree $0$. In phylogenetic trees the remaining {\em interior vertices} are unlabelled, of in-degree $1$ and of out-degree at least $2$~\citep{Steel2016}. In contrast to trees, we distinguish {\em tree vertices} with an in-degree of $1$ and an out-degree of at least $2$ from {\em reticulate vertices} with an in-degree of at least $2$ in phylogenetic networks. Each reticulate vertex has an out-degree of $1$. We do not allow vertices with an in- and out-degree of $1$~\citep{Huson2010}. Phylogenetic networks can contain pairs of parallel edges. We let $\mathds{N}(X)$ denote the set of rooted phylogenetic networks, and $\mathds{T}(X)$ be the set of rooted phylogenetic trees. 

\begin{figure}[!tb]
    \centering
    \includegraphics[width=0.5\linewidth, height=4cm]{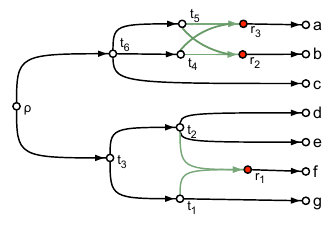}
    \caption{A rooted phylogenetic network $N$ on $X = \{a,b,c,d,e,f,g\}$ with the reticulate vertices labelled $r_*$ and coloured red, and the edges leading to them coloured green. The tree vertices are labelled $t_*$.}
    \label{fig:example-network}
\end{figure}

Let $N \in \mathds{N}(X)$ be any phylogenetic network. Following~\citet{Huson2010}, for any vertex $v \in V(N)$, except the root, the {\em lowest stable ancestor} (LSA) is defined as the lowest vertex $lsa_N(v) = lsa(v)$ that is part of all directed paths from the root to $v$ without being $v$ itself. Here, `lowest' refers to the vertex closest to the leaves of the network. Furthermore, the subscript $N$ indicates that $lsa(v)$ refers to vertices in the network $N$. For any tree vertex, the LSA equals its parent vertex~\citep{Huson2010}. In~\Cref{fig:example-network}, for example, $lsa_N(d)$ equals the vertex labelled $t_2$, whereas $lsa_N(r_1)$ is the vertex labelled $t_3$. The idea of the LSA can be extended by computing the LSA for multiple vertices. Assume that $Y \subseteq X$ denotes a non-empty set of taxa. The vertex $lsa(Y)$ is the lowest vertex in all directed paths from the root to every $y \in Y$. For short, we also write $lsa(a,b,c,\dots)$ instead of $lsa(\{a,b,c,\dots\})$. In~\Cref{fig:example-network}, for instance, $lsa_N(d,e,f,g)$ is the vertex labelled $t_3$.

Consider a network $N \in \mathds{N}(X)$ and a subset $Y \subseteq X$. The restricted network $N|Y$ is obtained from $N$ by restricting $N$ to the leaves in $Y$. The root of the new network $N|Y$ is $lsa_N(Y)$. After deleting all vertices and edges that are not part of any directed path from the new root to any $y \in Y$, all remaining vertices with an in- and out-degree of $1$ are suppressed. When $|Y| = 3$, the resulting network is called a {\em trinet}~\citep{Semple2021}. We say that a rooted tree with exactly three leaves is a {\em rooted triple}. In comparison to trinets, which can take an unlimited number of (unlabelled) shapes, there are only two distinct shapes for rooted triples~\citep{Steel2016}. 

Two phylogenetic networks $N, N' \in \mathds{N}(X)$ are isomorphic if a (directed graph) isomorphism exists between these networks, that is the identity map on $X$. We write $N \cong N'$~\citep{Steel2016}. The isomorphism of two trees can also be established by using the concept of rooted triples. The following result is classical and well known (for a proof, see, e.g. Theorem 1 from~\citet{Bryant1995}). \\

\begin{lemma} \label{lemmaIsomorphicTrees}
    Two phylogenetic trees $T,T' \in \mathds{T}(X)$ are isomorphic if and only if, for every subset $U \subseteq X$ such that $|U| = 3$, the trees $T$ restricted to $U$ and $T'$ restricted to $U$ are isomorphic:
    \begin{equation}
        T \cong T' \Longleftrightarrow T|U \cong T'|U ~ \forall U \subseteq X: |U|=3.
    \end{equation}
\end{lemma}

We also require the following notions. Let $\Sigma_X$ denote the group of permutations on $X$. Given a network $N \in \mathds{N}(X)$ and a permutation $\sigma \in \Sigma_X$, let $N^\sigma$ denote the network obtained by reordering the labels of the leaves according to $\sigma$.

Let $N \in \mathds{N}(X)$ be a phylogenetic network and $T \in \mathds{T}(X)$ be a phylogenetic tree.
We say that $T$ is {\em displayed} by $N$ if $T$ can be obtained from $N$ by the following process: For each reticulation vertex, delete all but one of its incoming arcs, then ignore any resulting leaves that are not in $X$ and suppress any vertices of in-degree and out-degree both equal to 1.

\section{Axiomatic properties for transformations that simplify phylogenetic networks}

Let $\mathds{N}'(X)$ be a defined subclass of the set of all phylogenetic networks $\mathds{N}(X)$ on any leaf set $X$.
We assume that the definition ensures the property 
that permuting the taxa of any member does not destroy membership.
Mainly, we focus on the case where $\mathds{N}'(X)=\mathds{T}(X)$ holds, and we also briefly discuss the case where $\mathds{N}'(X)$ is the class of normal networks $\tilde{\mathds{N}}(X)$. The transformations that we study are defined as follows: 
\begin{equation}
    \label{eq:phi_general}
    \varphi : \mathds{N}(X) \longrightarrow \mathds{N}'(X) \subseteq \mathds{N}(X),
\end{equation}
for every leaf set $X$ (including the subsets of any given leaf set).

We investigate the question of whether there are transformations that satisfy the following three specific properties (following the approach in~\citet{Dress2010}): \\
\newlist{contract}{enumerate}{3}
\setlist[contract]{label=$(P_{\arabic*})$}
\setlistdepth{3}
\begin{contract}
    \item $N \in \mathds{N}'(X) \Longrightarrow \varphi (N) = N$, 
    \item $\sigma \in \Sigma_X, N \in \mathds{N}(X) \Longrightarrow \varphi (N^\sigma) \cong \varphi (N)^\sigma $, 
    \item $Y \subseteq X, N \in \mathds{N}(X) \Longrightarrow \varphi (N|Y) \cong \varphi (N)|Y $. \\
\end{contract}

Property $P_1$ states that given a network that belongs to the subclass, the transformation returns the original network without further changes. It is an essential property because the transformation should not change the relationship of a set of taxa when the network correctly shows its evolutionary history~\citep{Dress2010}. This property implies that $N \in \mathds{N}'(X)$ if and only if $\varphi(N) \cong N$.

Property $P_2$ corresponds to a mathematical term that is referred to as {\em equivariance}~\citep{Bryant2017, Dress2010}. This means that the names of the taxa do not play any specific role in deciding how to simplify a network. In other words, when a transformation is applied to a network with permuted leaf labels, it should result in the same network as when we apply the transformation to the original network and then relabel the leaves. This property ensures that only the relationships between the taxa are relevant and not the way the  taxa are named or ordered~\citep{Dress2010}. Thus, this property could fail, for example, if  a transformation  depends on the order in which a user enters the species into a computer program, or if a  non-deterministic approach is applied in which  ties arising in some optimisation procedure are broken randomly. 

It is easily checked that all of the methods we consider in this paper satisfy these two properties. 

Property $P_3$ is more interesting and we refer to it as the {\em consistency} condition. It states that taking a subnetwork on a subset of taxa and applying the transformation gives the same network as that obtained by applying the transformation first and taking the subnetwork induced by the subset of taxa afterwards. The rationale for this axiom is as follows. Suppose a biologist adds new species to an existing network without changing the original network. Then, after transforming the new (enlarged) network, the network relationship between the original species should remain the same. It turns out that $P_3$ holds if and only if it holds for all subsets $U$ of $X$ of size $3$. We establish this result for transformations that convert a phylogenetic network into a phylogenetic tree.
To prove the statement, we first require another result. The proof of the following result is provided in the Appendix. \\

\begin{lemma} \label{lemmaSubsetNetworks}
    Let $N \in \mathds{N}(X)$ be a phylogenetic network and let $Y',Y$ with $Y'\subseteq Y \subseteq X$ denote two subsets. The network $N$ restricted to $Y$ and then restricted further to $Y'$ gives a network that is isomorphic to $N$ restricted to $Y'$. Formally: 
    \begin{equation} \label{eq:subset}
        Y' \subseteq Y \subseteq X, N \in \mathds{N}(X) \Longrightarrow (N|Y)|Y' \cong N|Y'.
    \end{equation}
\end{lemma}

\begin{lemma} \label{lemmaThirdProperty}
    Let $N \in \mathds{N}(X)$ be a rooted phylogenetic network and $\varphi$ denote a transformation, where $\mathds{N}'(X) = \mathds{T}(X)$. Property~$P_3$ holds if and only if~$P_3$ holds for subsets $U$ of size $3$:  
    \begin{equation} \label{eq:prop3}
        \varphi(N|Y) \cong \varphi(N)|Y ~ \forall Y \subseteq X \Longleftrightarrow \varphi(N|U) \cong \varphi(N)|U ~ \forall U \subseteq X: |U| = 3.
    \end{equation}
\end{lemma}

\begin{proof}
   Clearly, if ~$P_3$ is satisfied for all subsets $Y$ of $X$, then~$P_3$ holds for all $U \subseteq X$ of size $3$.
    
    We now prove that if~$P_3$ holds for all $U \subseteq X$ of size $3$, then~$P_3$ is satisfied for all $Y \subseteq X$. Assume that $P_3$ holds for all subsets $U$ of $X$ of size 3. Considering any $Y$, it suffices to show that 
        \begin{equation}
            \varphi (N|Y)|U \cong (\varphi (N)|Y)|U ~ \forall U \subseteq{Y}, |U| = 3,
        \end{equation}
    because of~\Cref{lemmaIsomorphicTrees}. By~\Cref{lemmaSubsetNetworks} we have:
        \begin{equation}
            (\varphi (N)|Y)|U \cong \varphi (N)|U.
        \end{equation}
   Thus, it remains to show that:
        \begin{equation}
            \varphi (N|Y)|U \cong \varphi (N)|U ~ \forall U \subseteq{Y}, |U| = 3.
        \end{equation}
    Let $N' = N|Y$ be the restricted network. We have:
        \begin{align}
            \varphi (N|Y)|U &= \varphi (N')|U &\text{ (substitution)} \nonumber \\
            &\cong \varphi (N'|U) &\text{ (assumption)} \nonumber \\
            &= \varphi ((N|Y)|U) &\text{ (substitution)} \nonumber \\
            &\cong \varphi(N|U) &\text{ (\Cref{lemmaIsomorphicTrees})} \nonumber \\
            &\cong \varphi(N)|U &\text{ (assumption)} \nonumber
        \end{align}
    Consequently, $P_3$ holds for all $Y \subseteq X$ if it holds for all $U \subseteq X: |U| = 3$, and vice versa.
\end{proof}

\section{Results}

We now consider various methods to transform an arbitrary rooted phylogenetic network into a rooted phylogenetic tree, and show that exactly one of these methods satisfies all three of the properties ($P_1, P_2, P_3).$

\subsection{Transformations that fail to satisfy the consistency condition}

\begin{figure}[!t]
    \begin{subfigure}{0.5\textwidth}
        \centering
        \includegraphics[width=0.9\linewidth, height=4cm]{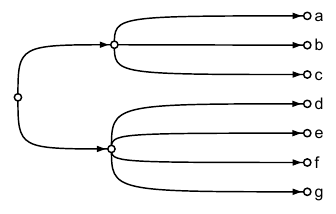} 
        \caption{$\varphi_{b}(N) = \varphi_{c}(N)$}
        \label{fig:example-network-blob-tree}
    \end{subfigure}
    \begin{subfigure}{0.5\textwidth}
        \centering
        \includegraphics[width=0.9\linewidth, height=4cm]{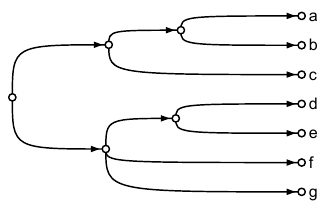} 
        \caption{$\varphi_{tc}(N)$}
        \label{fig:example-network-tight-clusters}
    \end{subfigure}
\caption{(a) The blob tree $\varphi_b(N)$ for the network $N$ in~\Cref{fig:example-network}, which is also the closed tree $\varphi_c(N)$. (b) The tree that corresponds to the tight clusters transformation $\varphi_{tc}(N)$.}
\label{fig:example-network_whole}
\end{figure}

We consider four transformations that transform a rooted phylogenetic network into a tree and satisfy $P_1$ and $P_2$. First, \citet{Gusfield2014} introduced the {\em blob tree} transformation, which we denote by $\varphi_b$.  In this transformation, a {\em biconnected component}\footnote{A biconnected component of a phylogenetic network is a maximal subgraph that cannot be disconnected by removing a single vertex (\cite{van10}).} is replaced by a new vertex; moreover, if  several biconnected components share a vertex, they are only replaced by one vertex (for further details, see~\citet{Gusfield2014}). An example blob tree for the network $N$ in~\Cref{fig:example-network} is shown in~\Cref{fig:example-network-blob-tree}.

Second, we include the {\em closed tree} transformation defined by~\citet{Huber2019}. For this method, we extract all closed sets of the network. For $N \in \mathds{N}(X)$, a  subset $Y \subseteq X$ is a {\em closed set}  of $N$  if $|Y| = 1$, or if  $|Y| \geq 2$ and the set of leaves of $N$ that are descended from the vertex $lsa_N(Y)$ is equal to $Y$.  In~\Cref{fig:example-network} the subset $Y = \{d,e,f,g\}$ corresponds to a closed set because the descending vertices of $lsa_N(Y)$ equal $\{d,e,f,g\}$.  From ~\citep{Huber2019}, the collection of all closed sets of a network forms a {\em hierarchy} (i.e. a collection of nested {\em clusters} (subsets of the leaf set)), and so corresponds to a tree (see~\Cref{fig:example-network-blob-tree}). We denote this transformation (from $\mathds{N}(X)$ to $\mathds{T}(X)$) as $\varphi_c$.

Another method is the {\em tight clusters} transformation described by~\citet{Dress2010}, which we denote by $\varphi_{tc}$. It works exactly as the previous transformation, except that we extract all `tight clusters' from the network instead of the closed sets.
Formally, for $N \in \mathds{N}(X)$, a non-empty subset $C$ of $X$ is a {\em tight cluster} of $N$ if there is a subset $S$ of vertices of $N$ for which $C$ is the set of leaves of $N$ reachable from $S$ and deleting $S$ from $N$ separates $C$ from $X\setminus C$. In~\Cref{fig:example-network}, for instance, $C = \{a,b\}$ is a tight cluster as $S = \{t_4, t_5\}$ is a valid subset that separates $\{a,b\}$ from $\{c,d,e,f,g\}$. The tight clusters of $N$ also form a tree~\citep{Dress2010}, which is shown in~\Cref{fig:example-network-tight-clusters}.

Finally, it is possible to apply any consensus method (e.g. strict consensus, majority rule, Adams consensus) on the set of trees $t(N)$ that are displayed by a network $N$ to get a transformation that satisfies $P_1$ and $P_2$. We considered the Adams consensus method $Ad(\{T_1, T_2, ...\})$ for a set of trees $\{T_1, T_2, ...\}$, which is a mathematically natural method based on the notion of `nesting'~\citep{Adams1972}.
To  construct $Ad(\{T_1, T_2, \ldots, T_k\})$ represent  each tree $T_i$ as a hierarchy. The maximal clusters of $Ad(\{T_1, T_2, \ldots T_k\})$ are simply the collection of non-empty intersections of the maximal clusters of $T_1, T_2, \ldots, T_k$ (these non-empty intersections partition $X$, and can be computed in polynomial time). 
The method then adds further clusters recursively by restricting the trees $T_1, T_2,\ldots,T_k$ to each block of the partition  of $X$ described. This process continues until the singleton sets $\{x\}$ (for all $x\in X$) are present in the resulting partitions, in which case the resulting hierarchy of sets corresponds to a rooted phylogenetic tree on leaf set $X$, which is $Ad(\{T_1, T_2, \ldots, T_k\})$. For further details, see \cite{Steel2016}. 

Our new {\em Adams consensus tree} transformation is defined as follows:
\begin{equation}
\label{ad1}
    \varphi_{ad}: \mathds{N}(X) \rightarrow \mathds{T}(X), ~ N \mapsto Ad(t(N)).
\end{equation}

It turns out that all four of the  transformations mentioned above fail to satisfy $P_3$. Stated formally: \\

\begin{proposition} \label{PropFail}
    The transformations $\varphi_b, \varphi_c, \varphi_{tc}$ and $\varphi_{ad}$ fail to satisfy $P_3$. \\
\end{proposition} 

Counterexamples for the proof of this result are provided in the second part of the Appendix.

\subsection{A transformation that satisfies the consistency condition}

We now describe a transformation that {\em does} satisfy all three of the properties ($P_1, P_2, P_3)$. First, we define the transformation. Then, we prove that the consistency condition ($P_3$) is satisfied (properties $P_1$ and $P_2$ are easily seen to also hold for this transformation). 

For any given rooted phylogenetic network $N$, one can compute a
rooted phylogenetic tree as follows:
First, for every reticulation vertex $r$ in $N$,
determine its lowest stable ancestor $lsa(r)$. Then, delete all edges leading into $r$
and create a new edge from $lsa(r)$ to $r$.
Second, repeatedly delete all unlabelled leaves
and suppress all vertices that have both in- and out-degree of $1$,
until no further such operation is possible~\citep{Huson2010}.
This tree $T_{lsa}(N)$ is uniquely defined and is called the LSA tree for $N$.

We denote this transformation as follows:
\begin{equation}
    \varphi_{lsa}: \mathds{N}(X) \rightarrow \mathds{T}(X), ~ N \mapsto T_{lsa}(N).
\end{equation}

\begin{figure}[!t]
    \centering
    \begin{subfigure}[b]{0.49\textwidth}
        \includegraphics[width=\textwidth]{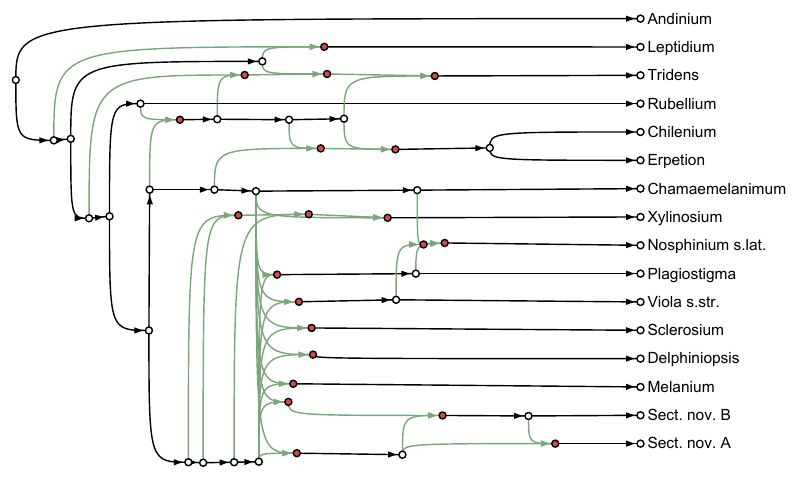}
        \caption{$N$}
        \label{fig:real-network}
    \end{subfigure}
    \begin{subfigure}[b]{0.49\textwidth}
        \includegraphics[width=\textwidth]{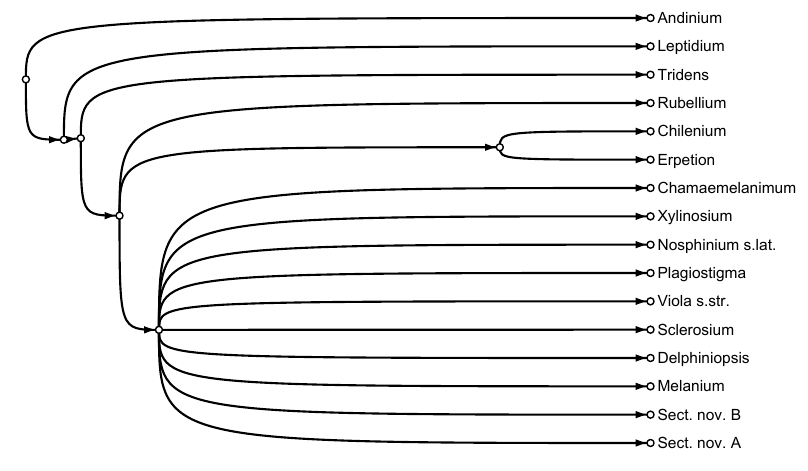}
        \caption{$\varphi_{lsa}(N)$}
        \label{fig:LSAtree}
    \end{subfigure}
    \caption{(a) A rooted phylogenetic network $N$ from~\citet{Jetten2018} based on a study from~\citet{Marcussen2014} with the reticulate vertices coloured red, and the edges leading to them coloured green. (b) The LSA tree $\varphi_{lsa}(N)$ of $N$.}
    \label{fig:real-data}
\end{figure}

\Cref{fig:real-data} shows the computation of the LSA tree for a biological network, which includes reticulate evolution, as investigated in the study of the {\em  Viola} genus by~\citet{Marcussen2014}. The network $N$ was studied in~\citet{Jetten2018}. We used {\em PhyloSketch}~\citep{Huson2020} to produce the LSA tree $\varphi_{lsa}(N)$.  

Before we prove that $P_3$ is satisfied, we look at a certain relationship between the LSA of a subset $Y$ and the LSA of two vertices $a,b \in Y$. \\

\begin{lemma} \label{lemmaLsaPartOfPath}
    Let $N \in \mathds{N}(X)$ be a phylogenetic network, and $Y \subseteq X$ denote a non-empty set with $a,b \in Y$. Then, $lsa_N(a,b)$ lies on all directed paths from $lsa_N(Y)$ to both $a$ and $b$. 
\end{lemma}

\begin{proof}
    First, we will derive a contradiction to show that $lsa_N(a,b)$ is either a descendant of $lsa_N(Y)$ or is equal to it. Assume that $lsa_N(Y)$ is a descendant of $lsa_N(a,b)$. The LSA of $Y$ is defined as the lowest vertex that is part of all directed paths from the root to each $y \in Y$. Therefore, the LSA of $a$ and $b$ is either $lsa_N(Y)$ itself or a descendant of the LSA of $Y$. This is a contradiction to the assumption. On the other hand, suppose that $lsa_N(Y)$ and $lsa_N(a,b)$ are different, and neither is a descendant of the other. This is impossible because $a,b \in Y$.

    Second, we show that $lsa_N(a,b)$ lies on all directed paths from $lsa_N(Y)$ to $a$ and $b$. We consider two cases. In the first case, suppose that $lsa_N(a,b) = lsa_N(Y)$. In this case, $lsa_N(a,b)$ is clearly part of every directed path from $lsa_N(Y)$ to any $y \in Y$. In the other case, $lsa_N(a,b)$ is a descendant of $lsa_N(Y)$. Without loss of generality, assume that at least one directed path from $lsa_N(Y)$ to $a$ does not include $lsa_N(a,b)$. According to the definition of the LSA, $lsa_N(Y)$ is part of all directed paths from the root to $a$,
    and this is also true for $lsa_N(a,b)$. If $lsa_N(Y)$ and $lsa_N(a,b)$ lie on  every directed path from the root to $a$, and $lsa_N(a,b)$ is a descendant of $lsa_N(Y)$, $lsa_N(a,b)$ needs to be part of all directed paths from $lsa_N(Y)$ to $a$, a contradiction to the assumption. The same applies for $b$. 
\end{proof}

There is a similar relationship when it comes to the LSA of reticulate vertices. \\

\begin{lemma} \label{lemmaLsaDescendant}
    Let $N \in \mathds{N}(X)$ be a phylogenetic network with $a,b \in X$. Consider a reticulate vertex $r \in V(N)$ that is part of a directed path from $lsa_N(a,b)$ to $a$. The LSA of $r$ is either a descendant of $lsa_N(a,b)$ or $lsa_N(a,b) = lsa_N(r)$ holds. 
\end{lemma}

\begin{proof}
    The two vertices $lsa_N(a,b)$ and $lsa_N(r)$ have to be comparable because $r$ lies on a directed path from $lsa_N(a,b)$ to $a$. Thus, suppose that $lsa_N(a,b)$ is a descendant of the LSA of $r$. Because $r$ is part of a directed path from $lsa_N(a,b)$ to $a$, there are at least two directed paths from the root to $r$, and thus from the root to $a$. All these directed paths need to include the $lsa_N(a,b)$ and $lsa_N(r)$ according to the definition of the LSA. Therefore, the $lsa_N(r)$ must either be $lsa_N(a,b)$ itself or a descendant of $lsa_N(a,b)$, which is a contradiction to the assumption.
\end{proof}

In contrast to all the other above mentioned transformations, the LSA tree transformation satisfies~$P_3$, as we now state formally. \\

\begin{theorem} \label{mainTheorem}
    The LSA tree method ($\varphi_{lsa}$) satisfies~$P_3$:  
    \begin{equation}
        (P_3) ~ Y \subseteq X, N \in \mathds{N}(X) \Longrightarrow \varphi_{lsa} (N|Y) \cong \varphi_{lsa} (N)|Y. \nonumber
    \end{equation}
    In other words, the LSA tree of a subnetwork on a subset of taxa equals the LSA tree of the original network restricted to the same subset of taxa. \\
\end{theorem}

In order to prove this theorem, we first require two results. \\

\begin{lemma} \label{lemmaSameLsaRestricted}
    Let $N \in \mathds{N}(X)$ denote a rooted phylogenetic network and let $Y$ be a non-empty subset of $X$. For every pair of vertices $a,b \in Y$, $lsa_N(a,b) = lsa_{N|Y}(a,b)$ holds. In other words, restricting the network does not affect the LSA of pairs of vertices.
\end{lemma}

\begin{proof}
    The root of the restricted network $N|Y$ corresponds to $lsa_N(Y)$. According to~\Cref{lemmaLsaPartOfPath}, the LSA of $a$ and $b$ is part of the directed path from $lsa_N(Y)$ to $a$ and $b$. All vertices that are part of a directed path from the new root to both $a$ and $b$ remain. Vertices that have an in- and out-degree of $1$ after deleting all vertices and edges that do not belong to the restricted network are suppressed. The vertex $lsa_N(a,b)$ has out-degree of at least $2$ because there are at least two directed paths (one to $a$ and another to $b$) going through this vertex. This means that $lsa_N(a,b)$ is not suppressed. Thus, $lsa_N(a,b)$ is equal to $lsa_{N|Y}(a,b) ~ \forall a,b \in Y$, as the directed path of the remaining leaves does not change. 
\end{proof}

\begin{lemma} \label{lemmaLsaReticulateNode}
    Let $T \in \mathds{T}(X)$ be the LSA tree of $N$. For all pairs of vertices $a,b \in X$, $lsa_N(a,b)$ equals $lsa_T(a,b)$.
\end{lemma}

\begin{proof}
    Two cases need to be distinguished.
    
    In the first case, no reticulate vertex lies on any directed path from $lsa_N(a,b)$ to both $a$ and $b$. By applying the LSA tree transformation, vertices with in- and out-degree of $1$ may be suppressed within these directed paths, but this does not affect the LSA of $a$ and $b$. Thus, $lsa_N(a,b) = lsa_T(a,b)$.
    
    In the second case, assume that there is at least one reticulate vertex $r$ in at least one of the directed paths (either from $lsa_N(a,b)$ to $a$ or $lsa_N(a,b)$ to $b$). This directed path changes by applying the LSA tree method as there is a new edge from $lsa_N(r)$ to $r$. By~\Cref{lemmaLsaDescendant}, either $lsa_N(a,b) = lsa_N(r)$ holds, or the LSA of $r$ is a descendant of $lsa_N(a,b)$. Therefore, the modification of the directed path does not affect the LSA of $a$ and $b$. Thus, $lsa_N(a,b) = lsa_T(a,b)$.
\end{proof}

\begin{proof}[Proof of~\Cref{mainTheorem}]
    It is sufficient to prove the following statement because of~\Cref{lemmaThirdProperty}: 
    \begin{equation}
        \varphi_{lsa}(N|U) \cong \varphi_{lsa}(N)|U ~ \forall U \subseteq X :|U| = 3.
    \end{equation}
    By definition of the LSA tree method, when $|U| = 3$, both $\varphi_{lsa}(N|U)$ and $\varphi_{lsa}(N)|U$ are rooted phylogenetic trees with exactly three leaves. These trees can only take two different shapes; they either have one or two interior vertices. Assume that $U = \{a,b,c\} \subseteq X$. We distinguish two different cases.
    
    In the first case, we apply the LSA tree method to $N|U$. In the resulting tree $a$ and $b$ are more closely related to each other than to $c$, without loss of generality. Therefore, $lsa_{\varphi_{lsa}(N|U)}(a,b) \neq lsa_{\varphi_{lsa}(N|U)}(a,c) = lsa_{\varphi_{lsa}(N|U)}(b,c)$. This implies:
    \begin{align}
        & lsa_{N|U}(a,b) \neq lsa_{N|U}(a,c) = lsa_{N|U}(b,c) &\text{(\Cref{lemmaLsaReticulateNode})} \nonumber \\
        \Rightarrow ~ &lsa_N(a,b) \neq lsa_N(a,c) = lsa_N(b,c) &\text{(\Cref{lemmaSameLsaRestricted})} \nonumber \\
        \Rightarrow ~ &lsa_{\varphi_{lsa}(N)}(a,b) \neq lsa_{\varphi_{lsa}(N)}(a,c) = lsa_{\varphi_{lsa}(N)}(b,c) &\text{(\Cref{lemmaLsaReticulateNode})} \nonumber \\
        \Rightarrow ~ &lsa_{\varphi_{lsa}(N)|U}(a,b) \neq lsa_{\varphi_{lsa}(N)|U}(a,c) = lsa_{\varphi_{lsa}(N)|U}(b,c). &\text{(\Cref{lemmaSameLsaRestricted})} \nonumber
    \end{align}
    Thus, $\varphi_{lsa}(N|U)$ is isomorphic to $\varphi_{lsa}(N)|U$ as the LSA of all pairs of leaves is the same in both trees, and as noted above there are only two possible shapes (having one or two interior vertices).
    
    In the other case, we consider the tree where all three leaves are equally related to each other. Thus, $lsa_{\varphi_{lsa}(N|U)}(a,b) = lsa_{\varphi_{lsa}(N|U)}(a,c) = lsa_{\varphi_{lsa}(N|U)}(b,c)$ holds. This implies:
    \begin{align}
        & lsa_{N|U}(a,b) = lsa_{N|U}(a,c) = lsa_{N|U}(b,c) &\text{(\Cref{lemmaLsaReticulateNode})} \nonumber \\
        &\Rightarrow lsa_N(a,b) = lsa_N(a,c) = lsa_N(b,c) &\text{(\Cref{lemmaSameLsaRestricted})} \nonumber \\
        &\Rightarrow lsa_{\varphi_{lsa}(N)}(a,b) = lsa_{\varphi_{lsa}(N)}(a,c) = lsa_{\varphi_{lsa}(N)}(b,c) &\text{(\Cref{lemmaLsaReticulateNode})} \nonumber \\
        &\Rightarrow lsa_{\varphi_{lsa}(N)|U}(a,b) = lsa_{\varphi_{lsa}(N)|U}(a,c) = lsa_{\varphi_{lsa}(N)|U}(b,c). &\text{(\Cref{lemmaSameLsaRestricted})} \nonumber 
    \end{align}
    Therefore, again, $\varphi_{lsa}(N|U)$ is  isomorphic to $\varphi_{lsa}(N)|U$. 
\end{proof}

\section{Concluding comments} 

In this paper, we have discovered that the LSA tree transformation satisfies all three desirable properties ($P_1$--$P_3$), whereas several other published transformations fail to satisfy $P_3$. 

Our results  suggest further questions and lines of inquiry. For example, is the LSA tree the only transformation from networks to trees that satisfies  $P_1$--$P_3$ and if not, can one classify the set of such transformations?

Secondly, for the transformation $N \mapsto Ad(t(N))$ described in (\ref{ad1}), is it possible to compute the $Ad(t(N))$ efficiently (i.e., in polynomial time in $|X|$)? Although computing the Adams consensus of a fixed set of trees can be done in polynomial time, the set $t(N)$ can grow exponentially with the number of reticulation vertices in $N$. More generally, for other consensus tree methods (e.g., strict consensus, majority rule) one can similarly define a transformation from phylogenetic networks to trees based on $t(N)$, so the same question of computational complexity  arises for these methods.

\begin{figure}[!t]
    \begin{subfigure}{\textwidth}
        \centering
        \includegraphics[width=0.7\textwidth, height=6cm]{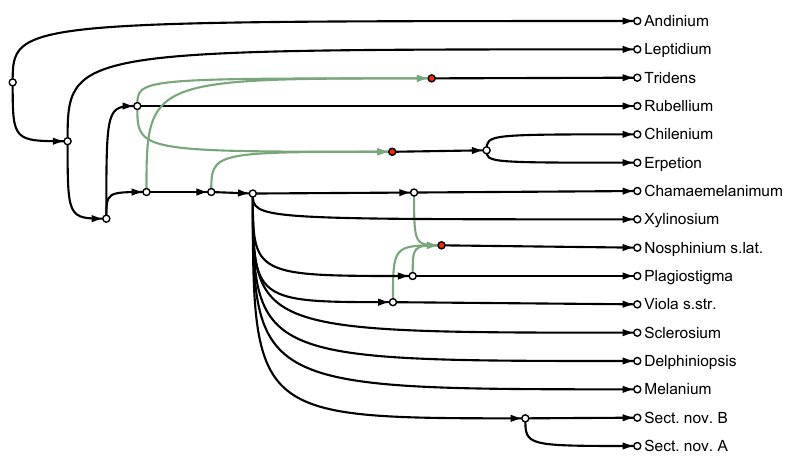}
        \caption{$\varphi_{no}(N)$}
        \label{fig:real-normal-network}
    \end{subfigure}
    \begin{subfigure}{0.5\textwidth}
        \centering
        \includegraphics[width=0.7\linewidth, height=2cm]{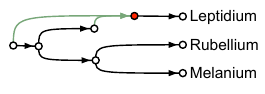} 
        \caption{$\varphi_{no}(N|Y)$}
        \label{fig:real-network-restricted}
    \end{subfigure}
    \begin{subfigure}{0.5\textwidth}
        \centering
        \includegraphics[width=0.7\linewidth, height=2cm]{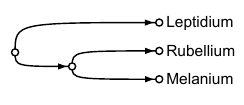} 
        \caption{$\varphi_{no}(N)|Y$}
        \label{fig:normal-network-restricted}
    \end{subfigure}
\caption{(a) The normalisation $\varphi_{no}(N)$ of the network $N$ in~\Cref{fig:real-network}. (b) The normalisation of the restricted network $\varphi_{no}(N|Y)$ with $Y~=~\{\text{Leptidium, Rubellium, Melanium}\}$. (c) The restricted normalisation $\varphi_{no}(N)|Y$ with $Y = \{\text{Leptidium, Rubellium, Melanium}\}$.}
\label{fig:counterexample_no}
\end{figure}

Finally, an alternative to transforming arbitrary phylogenetic networks to rooted trees is to consider transformations to other well-behaved network classes that allow for limited reticulation. A particular class of interest is the set of {\em normal} networks (i.e., every vertex is either a leaf or has at least one child that is a tree vertex, and there is no arc $(u,v)$ for which there is another path from $u$ to $v$) which have attractive mathematical and computational characteristics~\citep{Francis2021, Francis2021a, Kong2022}. 
Thus, it is of interest to consider  transformations from arbitrary phylogenetic networks on $X$ to the class of normal networks on $X$ and ask if such transformations can satisfy $P_1$, $P_2$ and $P_3$.
For the simple normalisation method $\varphi_{no}$ introduced in \citet{Francis2021},  $P_1$ and $P_2$ hold but $P_3$ fails. A counterexample is shown in~\Cref{fig:counterexample_no}. Whether or not there is a transformation that satisfies all three properties is thus an interesting question. One possible candidate could be the transformation described by~\citet{Willson2022}, but we do not consider this further in this paper.

\backmatter

\bmhead{Acknowledgements}

MS thanks the NZ Marsden Fund for funding support (23-UOC-003). {We also thank the two anonymous reviewers for their helpful comments on an earlier version of this manuscript.

\section*{Declarations}

The authors have no conflicts of interest to declare. There is no data associated with this paper. 

\section*{Appendix}

We first prove~\Cref{lemmaSubsetNetworks}. 

\begin{proof}
    Let $N \in \mathds{N}(X)$ be any network. Instead of the two subsets $Y$ and $Y'$, we consider $S = X \setminus Y$ and $S' = Y \setminus Y'$. Furthermore, let $N|Y = N^{-S} \in \mathds{N}(X\setminus S)$ denote the network restricted to $Y$. First,~\Cref{eq:subset1} is proven. We then establish~\Cref{eq:subset} by induction:
        \begin{equation} \label{eq:subset1}
            (N^{-x})^{-S'} \cong N^{-(\{x\} \cup S')} \text{ with } x \in X, x \notin S'.
        \end{equation} 
    To obtain $N^{-x}$ from $N$, we keep all vertices and edges that are part of any directed path from $lsa_N(X\setminus x)$ to any $n \in X\setminus x$. To get $(N^{-x})^{-S'}$, we also delete all vertices and edges that lie on any directed path from $lsa_N(X\setminus S')$ to any $s \in S'$, excluding the ones that are part of a directed path from $lsa_N(X\setminus S')$ to any $n \in X\setminus S'$.
    
    To obtain $N^{-(\{x\} \cup S')}$ from $N$, we can proceed the same way as described above because $\{x\} \cap S' = \emptyset$, by definition. First, delete $x$. Second, remove $S'$. Thus, $(N^{-x})^{-S'} \cong N^{-(\{x\} \cup S')}$. \Cref{eq:subset} can be rephrased as follows: 
        \begin{equation} \label{eq:subsetSame}
            (N^{-S})^{-S'} \cong N^{-(S \cup S')}.
        \end{equation}
    We prove~\Cref{eq:subsetSame} by induction on $k = |S|$. To start the induction, suppose $k = 1$. This case equals~\Cref{eq:subset1} and was proven above. Now, assume that~\Cref{eq:subsetSame} is true for some $k$. We show that the statement then holds for $k+1$. Suppose that $S = \{s_1, ..., s_k, s_{k+1}\}$. We have:
    \begin{align}
        (N^{-S})^{-S'} &= (N^{-\{s_1,...,s_k,s_{k+1}\}})^{-S'} &\text{(substitution)} \nonumber \\
        &= (N^{-(\{s_1,...,s_k\} \cup \{s_{k+1}\})})^{-S'} &\text{(set operation)} \nonumber \\
        &\cong ((N^{-\{s_1,...,s_k\}})^{-s_{k+1}})^{-S'} &\text{(induction hypothesis)} \nonumber \\
        &\cong (N^{-\{s_1,...,s_k\}})^{-(\{s_{k+1}\} \cup S')} &\text{(\Cref{eq:subset1})} \nonumber \\
        &\cong N^{-(\{s_1,...,s_k\} \cup \{s_{k+1}\} \cup S')} &\text{(induction hypothesis)} \nonumber \\
        &= N^{-(S \cup S')} &\text{(substitution)} \nonumber
    \end{align}
    \Cref{eq:subsetSame} is proven, and thereby, \Cref{eq:subset}. 
\end{proof}

\begin{figure}[!t]
    \begin{subfigure}{0.5\textwidth}
        \centering
        \includegraphics[width=0.7\linewidth, height=3.5cm]{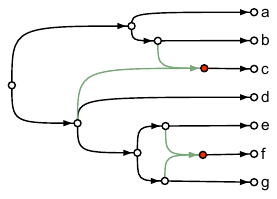}
        \caption{$N$}
        \label{fig:network}
    \end{subfigure}
    \begin{subfigure}{0.5\textwidth}
        \centering
        \includegraphics[width=0.7\linewidth, height=3.5cm]{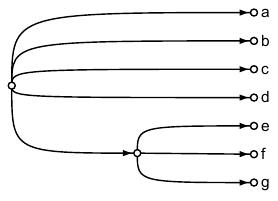}
        \caption{$T = \varphi_b(N) = \varphi_c(N) = \varphi_{tc}(N)$}
        \label{fig:tree}
    \end{subfigure}
    \begin{subfigure}{0.5\textwidth}
        \centering
        \includegraphics[width=0.7\linewidth, height=3cm]{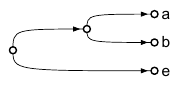} 
        \caption{$\varphi_{b}(N|\{a,b,e\}) = \varphi_{c}(N|\{a,b,e\}) = \varphi_{tc}(N|\{a,b,e\})$}
        \label{fig:network-restricted}
    \end{subfigure}
    \begin{subfigure}{0.5\textwidth}
        \centering
        \includegraphics[width=0.7\linewidth, height=3cm]{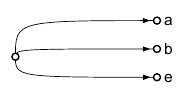} 
        \caption{$T|\{a,b,e\}$}
        \label{fig:tree-restricted}
    \end{subfigure}
\caption{(a) A rooted phylogenetic network $N \in \mathds{N}(X)$ on leaf set $X~=~\{a,b,c,d,e,f,g\}$. (b) The blob tree $\varphi_b(N)$, the closed tree $\varphi_c(N)$ and the result of the tight clusters transformation $\varphi_{tc}(N)$ give rise to the same tree $T$. (c) The restricted network $N|\{a,b,e\}$, which corresponds to its transformation. (d) The restricted tree $T|\{a,b,e\}$.}
\label{fig:counterexample_b_c_ct}
\end{figure}

\begin{figure}[!t]
    \begin{subfigure}{0.5\textwidth}
        \centering
        \includegraphics[width=0.7\linewidth, height=3.5cm]{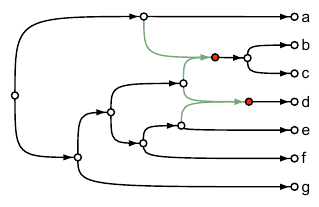}
        \caption{$N$}
        \label{fig:network-a}
    \end{subfigure}
    \begin{subfigure}{0.5\textwidth}
        \centering
        \includegraphics[width=0.7\linewidth, height=3.5cm]{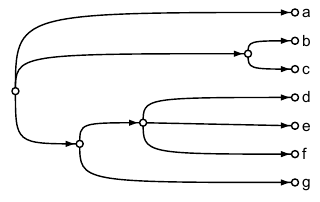}
        \caption{$\varphi_{ad}(N)$}
        \label{fig:tree-a}
    \end{subfigure}
    \begin{subfigure}{0.5\textwidth}
        \centering
        \includegraphics[width=0.7\linewidth, height=3cm]{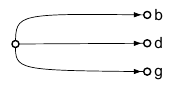} 
        \caption{$\varphi_{ad}(N|\{b,d,g\})$}
        \label{fig:network-restricted-a}
    \end{subfigure}
    \begin{subfigure}{0.5\textwidth}
        \centering
        \includegraphics[width=0.7\linewidth, height=3cm]{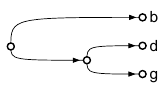} 
        \caption{$\varphi_{ad}(N)|\{b,d,g\}$}
        \label{fig:tree-restricted-a}
    \end{subfigure}
\caption{(a) A rooted phylogenetic network $N \in \mathds{N}(X)$ on leaf set $X~=~\{a,b,c,d,e,f,g\}$. (b) The Adams consensus tree $\varphi_{ad}(N)$ of $N$. (c) The restricted transformed network $\varphi_{ad}(N|\{b,d,g\})$. (d) The restricted Adams consensus tree $\varphi_{ad}(N)|\{b,d,g\}$.}
\label{fig:counterexample_ad}
\end{figure}

Secondly, turning to the proof of~\Cref{PropFail}, \Cref{fig:counterexample_b_c_ct} provides a counterexample for the transformations $\varphi_b, \varphi_c$ and $\varphi_{tc}$. The counterexample for the transformation $\varphi_{ad}$ is shown in~\Cref{fig:counterexample_ad}. 

\newpage 

\bibliography{HEISS_et_al}

\end{document}